\documentstyle[11pt,arnps,epsf]{article}
\newcommand{\cO}{{\cal O}}

\newcommand{\half}{\mbox{\small $\frac{1}{2}$}}
\def\imag{\mathop{\rm Im}}
\def\lsim{\;\raisebox{-.4ex}{\rlap{$\sim$}} \raisebox{.4ex}{$<$}\;}
\def\MSbar{{\overline{\rm MS}}}

\def\real{\mathop{\rm Re}}

\preprint{FERMILAB-PUB-93/058-T}
\title{\bf PROGRESS IN QCD USING LATTICE GAUGE THEORY}
\author{ANDREAS S. KRONFELD and PAUL B. MACKENZIE \\[0.3cm]
{\sl Theoretical Physics Group, Fermi National Accelerator  Laboratory,}
        \\ {\sl P.O. Box 500, Batavia, IL 60510, USA}}

\date{March 1993}

\begin{document}

\maketitle
\vfil
\vfil
%
%%%%% \noindent {\rm SHORT TITLE: Lattice QCD}
\noindent to appear in {\em Annual Review of Nuclear and Particle Science}
\vfil
\pagebreak
\tableofcontents
% \pagebreak
\vspace{1.0cm}

\section{Introduction}
Quantum chromodynamics (QCD) is the only serious candidate for the theory
of strong interactions.
It is supported by overwhelming qualitative evidence and a growing
body of quantitative evidence.
Lattice gauge theory is the only fundamental formulation of QCD allowing
the calculation of all its consequences in both the high and low energy
regimes.

Low energy QCD is worth studying not only for its own sake, but also
for its role in understanding what lies beyond the standard model.
At present, the only experimental clues for this puzzle are the fundamental
parameters of the standard model.
Of these, the values of the strong coupling constant, all of the quark
masses except the top quark mass, and most of the
Cabibbo-Kobayashi-Maskawa (CKM) matrix elements
either are now or soon
will be dominated by theoretical uncertainties that can be attacked
with lattice QCD.
Table~\ref{table:sm} contains a list of the most fundamental quantities
in the standard model.
Where appropriate it also indicates how lattice QCD will play an
important role, and the section(s) of this article containing relevant
material.
The central theme of this review is standard model phenomenology, with
emphasis on lattice calculations needed to determine the
parameters and to understand the reliability of the determinations.

\begin{table}
\caption[table:sm]{Parameters of the standard model and lattice
calculations which will help determine them.
Ranges for CKM matrix elements assume unitarity but not three
generations.
Numerical values taken from Reference~\cite{PDG92}, except
$\sin\delta$ and $\theta_{\rm QCD}$.}\label{table:sm}
\begin{center} \begin{tabular}{c@{\hspace{2.0em}}c@{\hspace{2.0em}}cl}
\hline \hline
   parameter       & value or range &
related lattice calculations & section(s)  \\
\hline
  $\alpha_{\rm em}$  &   $1/137.036$    &         &                 \\
     $10^5G_F$       & 1.166 GeV$^{-2}$ &         &                 \\
$\alpha_\MSbar(M_Z)$ &   0.110--0.118   &
$\Delta m_{\mbox{\scriptsize 1P--1S}}$; scaling & \ref{onia}; \ref{alpha} \\
$m_Z$                &    91.17 GeV     &         &          \\
$m_H$                &   $>48$ GeV      &         &          \\
$m_e$                &   0.51100 MeV    &         &          \\
$m_\mu$              &    105.66 MeV    &         &          \\
$m_\tau$             &    1.78 GeV      &         &          \\
% $\hat{m}/m_s$        &   0.034--0.044   &
%   $m_{\pi}^2/m_K^2$  & \ref{spectrum}, \ref{quark-mass} \\
% $\Delta m^2_{du}/m^2_s$ &  0.021--0.025 &         &          \\
$m_u$                &     2--8   MeV   &
$m_\pi^2$, $m_K^2$ & \ref{spectrum}, \ref{quark-mass} \\
$m_d$                &     5--15  MeV   &
$m_\pi^2$, $m_K^2$ & \ref{spectrum}, \ref{quark-mass} \\
$m_s$                &   100--300 MeV   &
$m_K^2$ & \ref{spectrum}, \ref{quark-mass} \\
$m_c$                &   1.3--1.7 GeV   &
$m_{J/\psi}$   & \ref{onia}, \ref{quark-mass}\\
$m_b$                &   4.7--5.3 GeV   &
$m_{\Upsilon}$ & \ref{onia}, \ref{quark-mass}\\
$m_t$                &   $>91$ GeV      &         &          \\
$|V_{ud}|$           &      0.974       &         &          \\
$|V_{us}|$           &      0.220       &         &          \\
$|V_{ub}|$           &   0.002--0.007   &
$B\rightarrow\rho l\nu$ & \ref{semi-leptonic} \\
$|V_{cd}|$           &   0.179--0.228   &
$D\rightarrow\pi l\nu$  & \ref{semi-leptonic} \\
$|V_{cs}|$           &   0.864--0.975   &
$D\rightarrow K   l\nu$ & \ref{semi-leptonic} \\
$|V_{cb}|$           &   0.032--0.054   &
$B\rightarrow D   l\nu$ & \ref{semi-leptonic} \\
$|V_{td}|$           &   0.0--0.14      &
$f_B$, $B_B$; $B_K$  & \ref{decay-constant}, \ref{mixing} \\
$|V_{ts}|$           &   0.0--0.45      &
$f_{B_s}$, $B_{B_s}$ & \ref{decay-constant}, \ref{mixing} \\
$|V_{tb}|$           &   0.0--0.9995    &         &                 \\
$\sin \delta$        &    $\neq 0$      &
$B_K$, $B_B$, $B_{B_s}$  & \ref{mixing}    \\
$\theta_{\rm QCD}$   &   $<10^{-9}$     & $d_n$   &                 \\
\hline \hline
\end{tabular} \end{center}\end{table}

When lattice gauge theory was first introduced by Wilson in 1974
\cite{Wilson}, several calculational approaches were suggested,
including strong coupling expansions and various renormalization group
methods.
Monte Carlo methods were first applied to pure gauge theory
% by Creutz and by Wilson
in 1979 \cite{Wil79,Creutz}.
Methods for treating quarks in Monte Carlo calculations were introduced
% by Weingarten and by Parisi and collaborators
in 1981 \cite{Parisi}.
Although these initial calculations of the hadron spectrum
had approximately the reliability of
the nonrelativistic quark model, it was clear, at least in principle,
how to develop them into genuine first principles QCD calculations.
They initiated the wave of effort leading to the calculations described
in this article.

A very brief overview of lattice methods is given in Section~\ref{methods}.
That section also details the sources of error and uncertainty
which must be understood and eliminated as lattice methods
evolve into true first principles calculations.
%We pay particular attention to an
Most of the calculations we discuss employ an
approximation introduced in
References~\cite{Parisi}, called the ``quenched'' or ``valence''
approximation.
The former name is more common in the literature, but the latter one is,
perhaps, more descriptive:
the quenched approximation treats valence quarks exactly and ignores the
effects of sea quarks.
%Because the quenched approximation reduces the computational demands,
%the available computer time can be invested the analysis of the other
%uncertainties.
%Hence, calculations quenched lattice QCD are much more mature than in
%full QCD with sea quarks.

In Section~\ref{onia} we discuss lattice calculations of the the $\psi$
and $\Upsilon$ systems.
Solid error analysis is easiest to produce in these simple systems
because of the possibility of using nonrelativistic methods.
The 1P--1S splitting, $\Delta m_{\mbox{\scriptsize 1P--1S}}$, in these
systems is insensitive to the most serious sources of error in
lattice calculations.
This makes it the ideal quantity for setting the scale, i.e.\ converting
from lattice to physical units.
The calculation of the
 light hadron spectrum is one of the original goals of lattice gauge
theory, and a completely reliable calculation is still a major piece of
unfinished business.
For many years the progress was incremental.
As discussed in Section~\ref{spectrum}, however, recent developments may
represent a new standard in the thoroughness in the treatment of errors.
%of the error analysis,
%yielding results in the quenched approximation in remarkable agreement
%with experiment.

Section~\ref{QCD} discusses applications of the spectrum calculations
towards determining the fundamental parameters of QCD, the quark masses
and strong coupling constant.
The lattice determination of the latter from the $\psi$ and $\Upsilon$ spectra
is already competitive with
perturbative determinations from high-energy scattering experiments.
As with the ``traditional'' results for $\alpha_S$, there is still
a phenomenological component in the lattice determinations.
The major
uncertainty in the lattice QCD results come from modeling the effects of
sea quarks.
However, in lattice QCD the path is clear towards eliminating the
modeling completely.
Hence, in the long run, the most precise determination of $\alpha_S$
will likely come from lattice QCD.

Although the spectrum calculations are indisputably essential to the
verification of QCD,
%it has become clear in the last decade that
many lattice
calculations in weak-interaction phenomenology are of even
greater importance to the standard model.
This is the subject of Section~\ref{WIP}.
The CKM matrix is responsible for (at least) four parameters, and
one would like to overdetermine it to test whether there are further
generations (with massive neutrinos).
The good news is that some of the associated hadronic physics, such as
the kaon ``$B$'' parameter,
can be calculated with comparable or greater
 reliability than the light hadron spectrum.
%comes especially to mind, and we predict that
%similarly reliable calculations of semi-leptonic form factors will
%become available in the near future.

Since an excellent introduction appeared in this series eight years ago
\cite{Has85}, this article does not review the foundations of lattice
field theory.
Another pedagogical introduction is in Reference~\cite{Kro92}.
For an encyclopedic overview of the activity in lattice field theory,
the reader can consult any recent proceedings of the annual
international symposium on lattice field theory \cite{LAT87}--\cite{LAT92}.

This review also omits several important applications of lattice field
theory.
The study of the deconfinement temperature in SU(3) gauge theory without
quarks was influential.
It was the first careful application of large scale Monte Carlo
methods to a quantity whose value was not well known in advance
\cite{Got85,Chr86}.
More recent work with three light quarks suggests that the structure of the
phase transition in QCD may depend sensitively on the mass of the
strange quark \cite{Chr92}.
For a review of these and other topics in QCD thermodynamics,
see Reference~\cite{Petersson}.
Analytical and numerical methods of lattice field theory have been used
to obtain upper bounds on the masses of the Higgs boson and of heavy
quarks in the standard model \cite{weak}.
Most proposals for strongly coupled models of electroweak symmetry
breaking require a lattice regularization for chiral fermions.
This problem is still unsolved; the status of of current ideas for
solutions is reviewed in Reference~\cite{Petcher}.

With one exception, all of the QCD entries in Table~\ref{table:sm} are
based on meson properties.
The bound on the strong CP-violating parameter $\theta_{\rm QCD}$,
however, comes from the neutron electric dipole moment
\cite{Bal79,Alt81}.
Lattice QCD calculations of such baryon properties are more difficult
than comparable ones for mesons, so there has been less systematic work.
Another QCD topic of vital interest is the study of glueballs and other
bound states that would not appear in the quark model.
Despite the algorithmic improvements of recent years \cite{Tep87,Ape88},
glueball mass calculations still suffer from a small signal-to-noise
ratio.
Therefore, it seems appropriate to postpone a review of baryon and
glueball phenomenology.

\clearpage

\section{Lattice Methodology} \label{methods}
Because of our emphasis on standard-model phenomenology, we
omit discussion of most technical details.
For a more thorough introduction, see References~\cite{Has85,Kro92}.
We provide here only a schematic overview of lattice methods,
plus a brief discussion for nonexperts
 of the most important sources of uncertainty in
lattice calculations.

\subsection{Methods}
The path integral formulation of quantum field theory is used to
define lattice QCD:
\begin{equation}
Z = \int{\cal D}U\ {\cal D}\psi\ {\cal D}\bar{\psi}\; e^{-S_G-S_Q}.
\end{equation}
The integration is over each $U$ variable (SU(3) matrices representing
the gluon fields, defined on each link of the lattice) and each $\psi$
and $\bar{\psi}$ field (anticommuting variables
representing quark fields, defined on each site of the lattice).
The standard action used in almost all lattice calculations is
\begin{equation}\label{plaquette-action}
S_G = \frac{\beta}{6} \sum_{x,\mu,\nu} P_{\mu\nu}(x)
\end{equation}
for the gluons.
The ``plaquette'' $ P_{\mu\nu}(x)$ is the trace of the product of the
$U$ matrices around the elementary square at $x$ in the $\mu$-$\nu$
plane.
There are two commonly used formulations of lattice fermions.
Except for quark masses and $B_K$, the calculations we discuss use the
Wilson formulation
\begin{eqnarray}\label{Wilson-fermion-action}
S_Q&=&-\kappa\sum_{x,\mu}
\bar{\psi}_x[(1-\gamma_\mu)U_{x,\mu}    \psi_{x+\hat{\mu}}
+(1+\gamma_\mu) U_{x-\hat{\mu},\mu}^\dagger   \psi_{x-\hat{\mu}}]\\
\nonumber
&+&\sum_x\bar{\psi}_x\psi_x.
\end{eqnarray}
Wilson fermions allow the proper number of flavors at the expense of a
difficult handling of chiral symmetry.
When chiral symmetry is crucial another formulation is available,
``staggered fermions,'' which maintain an exact chiral symmetry, but
then the number of flavors is a multiple of four.

The parameters in the lattice action are $\beta$ and the ``hopping
parameter'' $\kappa$.
The bare lattice coupling constant is given by $\beta=6/g_0^2$.
The bare quark mass is related to $\kappa$ by $\kappa=1/(8 + 2m_0 a)$.
Using these identifications it is easy to show that in the zero lattice
spacing limit, this action reduces to the usual QCD action,
$S_Q=\int d^4x\,\bar{\psi}(x)({\partial_\mu\gamma_\mu}+m_0)\psi(x)$,
and $S_G=(1/4g_0^2)\int d^4x(F_{\mu\nu}^a)^2$,
plus errors which vanish with the lattice spacing.
The lattice spacing $a$ is made to vanish by taking
$\beta\rightarrow\infty$, keeping physical quantities fixed.
Note, however, that much of the literature uses ``lattice units,'' where
$a=1$.

The path integral formalism shows how the correlation functions of
hadron operators $\Phi_t(U,\psi,\bar{\psi})$ (at time $t$) behave:
\begin{eqnarray}
\langle \Phi_t \Phi_0^\dagger \rangle 	\label{eq:hcorr}
  &=&  \frac{1}{Z} \int{\cal D}U\ {\cal D}\psi\ {\cal D}\bar{\psi}\;
  \Phi_t\Phi_0^\dagger e^{-S_G-S_F}	\\  \label{eq:higherstates}
  &=& \sum_{\beta} |\langle0|\hat{\Phi}|\beta \rangle|^2
e^{-tE_{\beta}}.
\end{eqnarray}
The hadronic correlation functions decay as sums of exponentials if the
theory is formulated in Euclidean space.
The rates of decay $E_\beta$ of these exponentials are the energies of the
states $\beta$.
The coefficients of the exponentials are related to hadronic matrix elements.
The Euclidean-space formulation has many numerical advantages.
For example, for $t$ large enough, it is possible to isolate the state
with the lowest energy.
However, the contributions to the correlation function of all the states
$\beta$, other than the lowest state, produce errors which must
estimated and eliminated.

In integrals such as Equation~\ref{eq:hcorr} the integration over the
Fermi fields can performed explicitly, leaving a gauge-field integral.
This step expresses the correlation function in terms of
quark propagators in a background gauge field.
The integration over the gauge fields is evaluated by Monte Carlo with
importance sampling, yielding ensembles of lattice gauge fields.
The gauge-field integral is approximated as a finite sum, introducing a
statistical error.
New gauge field configurations are generated from previous ones and
are correlated with them.
This effect must be carefully accounted for in the statistical analysis.

Quark propagators are solutions of the discrete Dirac equation,
which is a sparse matrix equation.
Sparse matrix methods are used to produce the propagators for each lattice.
These algorithms must be employed much more often in full QCD than in
the quenched approximation, to account for the back-reaction of the sea
quarks on the gluons.

\subsection{Error Analysis}
These sources of uncertainty in this section must be individually
understood if numerical lattice QCD is to become a widely accepted
calculational tool.
A first pass at a thorough enumeration has been attempted for only a few
of the simplest quantities, but there is a good hope that full error
analysis (in the quenched approximation) will be extended to many more
quantities over the next two or three years, including some extremely
interesting ones.
Therefore, we shall now discuss how the various sources of error can be
reduced:

\paragraph{Statistical errors.}
Any Monte Carlo procedure has statistical uncertainties.
In lattice QCD these may be the errors which are currently under best
control.
A subtlety is to cope with the correlations among subsequent
configurations.
These correlations can extend over stretches in the Monte Carlo chain,
especially for the algorithms used in full QCD.
Another subtlety is that the statistical uncertainty of quantities
calculated within a single ensemble of gauge fields are correlated.
Hence, ratios of similar quantities usually have smaller statistical
errors than the quantities themselves.

\paragraph{Finite lattice spacing errors.}
If the lattice action in is expanded in powers of the lattice
spacing, one obtains the standard continuum action of QCD, plus
an infinite series of unwanted, higher dimension operators whose effects
on masses (or other quantities derived from the spectrum) vanish as
powers of the lattice spacing.
Their effects can be systematically
eliminated by adding higher dimension correction
operators to the lattice action \cite{Symanzik}.
An order of magnitude estimate of their effects is $a \Lambda_{\rm QCD}$
to the appropriate power.
For $\beta=6.0$, $a^{-1} \approx 2$ GeV,
this is around 10--15\% for the simple $O(a)$ correction for Wilson
fermions, and 1--2\% for the more complicated $O(a^2)$ errors of the
quark and gluon actions.

The most serious of these errors, the $O(a)$ error for Wilson
fermions, can be corrected by the addition of a single term to
the fermion action~\cite{WOSH}
\begin{equation}~\label{eq:WOSH}
\delta S_Q= i g \frac{c}{2}\kappa \sum_{x,\mu,\nu}
\bar{\psi}_x \sigma_{\mu\nu} F_{\mu\nu} \psi_x,
\end{equation}
where $\sigma_{\mu\nu}= \frac{i}{2} [\gamma_\mu,\gamma_\nu]$,
the $\gamma_\mu$ are Euclidean gamma matrices, and $F_{\mu\nu}$ reduces
to the QCD field strength tensor as $a\rightarrow0$.
Direct calculational evidence of the importance of this correction
has been given in Reference~\cite{Martinelli} and in the charmonium
calculations described in Section~\ref{onia}.
A more careful examination reveals that the coefficient $c$ depends on
the bare coupling.
In perturbation theory $c=1+c_1g_0^2+\cdots$.
The systematic program of adding corrections like $\Delta S_Q$ and
calculating their coefficients is called ``improvement''
\cite{Symanzik}.

\paragraph{Finite volume errors.}
Numerical calculations of lattice QCD are done in a finite volume,
because then there is a finite number of degrees of freedom, which can
be stored in the finite memory of a computer.
Finite volume errors are nonperturbative properties of QCD, and thus more
complicated to analyze.
However, for periodic boundary conditions, they are expected to fall
exponentially with lattice size.
It is therefore a reasonable goal to increase the lattice until they
are really negligible.
The asymptotic errors are known for the proton mass \cite{Luescher},
and quite small for lattices of reasonable size.
The functional form in the intermediate region is unknown and must be
carefully determined by numerical calculation.
For an excellent technical review of the state of the art on this
and many other issues in hadron spectroscopy, see \cite{Ukawa}.

\paragraph{The effects of higher mass states.}
In extracting the properties of the ground state from  correlation functions
such as Equation~\ref{eq:higherstates}, the contamination from more massive
states with the same quantum numbers must be estimated and reduced.
This is most often done by separating the creation and destruction operators
far enough that only a single exponential of the sum in
Equation~\ref{eq:higherstates} is visible within the statistical errors.
This approach has the drawbacks that it is limited by increasing statistical
errors as the operators are separated, and that systematic uncertainty
estimates are difficult.
Another approach is to vary the operator or matrix of operators $\Phi$ to
maximize the overlap with the desired state and minimize the overlap
with the rest.

\paragraph{The extrapolation to physical quark mass.}
Current lattice algorithms for sparse matrix inversion (and thus for the
inclusion of the effects of sea quarks)
become much more computationally demanding,
and sometimes fail entirely, as the quark mass is reduced toward its physical
value.  Current calculations rarely go below $m_\pi/m_\rho\sim 0.4$,
compared to the physical value of 0.18.  Leading-order chiral behavior
is usually assumed in extrapolating to the physical quark mass
($m_\pi^2$ and the masses of the other hadrons proportional to $m_q$).
The size of deviations from linearity for mesons of nearly the mass of the
kaon are controversial among workers in chiral perturbation theory.
In lattice QCD they must be determined by numerical calculations.

\subsection{The quenched approximation}
While gradual and systematic programs exist for the elimination of the
above sources of error, no better way is known to improve on the
quenched approximation
than to include all effects of sea quarks at once.
Formulas for the effects of small numbers of internal quark loops may be
derived in terms of correlations of hadronic operators with the fermionic
effective action, but they appear to be even harder to handle than
the exact formula.
Algorithms for inclusion  of quark loops are much more
computationally demanding than those which omit them, so the analysis of
the other sources of uncertainty is much cruder for calculations which include
them.
This review will therefore concentrate on calculations which omit them.
In a few cases, but not in general, it is possible to make
phenomenological estimates of the accuracy of the quenched approximation.
We will return in Section~\ref{Summary}
 to the general case of the effects of sea quarks.

%\subsection{Perturbation theory}
%well-defined \\
%badly behaved because of tadpoles \\
%mean-field improves lowest order \\
%renormalized couplings improve higher orders.

\section{Establishing and Testing Lattice Methods}

\subsection{The $\psi$ and $\Upsilon$ Systems} \label{onia}

The discovery of charmonium, the bound states of $c$ and $\overline{c}$
quarks, with their clear positronium-like spectra, provided an important
psychological boost to the belief in the reality of quarks.
The success nonrelativistic potential models \cite{Eetal}
in accounting for these
spectra provided a boost to the acceptance of QCD as the theory of
strong interactions, since the  models became equivalent to leading
order QCD
 in a well defined
limit: the large quark mass limit.
The $ \psi$ and $\Upsilon$ systems are proving crucial in establishing
the accuracy of lattice calculations
because nonrelativistic reasoning opens ways of checking and rechecking
methods of error analysis that are unavailable for the lighter hadrons.
%They  deserved to be called the ``Hydrogen atom
%of QCD''.
% even though theoretical methods at the time of their discovery
%were not then powerful enough to solve them completely from first
%principles.

%Quarkonia have received little attention from lattice theorists
%until relatively recently.  This is partly due to the fact
%that their gross features can be well understood
%with the excellent phenomenology of potential models.
%They therefore seemed less exciting for lattice calculations since they
%were already understood on the basis of good QCD based phenomenology.
%However,
As Lepage \cite{ThackerLepage},
 has emphasized, now that lattice methods are coming into fruition,
it is these simple systems which will provide the best early tests of
lattice methods.
There are some technical reasons for this.  Since the quarks are heavy,
the extrapolation to the physical light quark mass required in light
hadron calculations  is unnecessary.
The propagators of heavy quarks are much quicker to calculate on the
lattice than those of light quarks.
Further, since the heavy mesons are smaller than the light hadrons,
smaller physical volumes suffice.
However, the most important fact making the properties of these mesons
the easiest to calculate on the lattice is the one that made
possible the good phenomenological treatment of them twenty years ago:
they are nonrelativistic systems.
This means means that potential models and
the nonrelativistic arguments justifying them can be used both to
guide the physics expectations of the lattice calculations,
and to  supplement the analysis of corrections and uncertainties in
the lattice calculations.

%Potential models were used immediately after the discovery of the
%$J/\psi$ system to calculate many properties of the system \cite{Eetal}.
%These included the spin averaged level splittings, leptonic widths,
%E1 transitions, and later hyperfine splittings.
%The first calculations of the Cornell group used a pure Coulomb plus
%linear potential, without asymptotic freedom.
%Asymptotically free Coulomb plus linear potentials, introduced by
%Richardson, produced improved agreement with data \cite{Richardsonpot}.
%In particular, the wave function at short distances and therefore
%the leptonic width ware significantly reduced.
%%This will be relevant to quenched lattice calculations.
%Some of the classic applications of potential models are among the most
%straightforward bread and butter applications of lattice methods.
%These include the 1P--1S splitting, and the hyperfine splitting
%and leptonic width of the 1S.

Potential models play an important role in defining physics expectations
for lattice charmonium calculations.
For example, the part of the hyperfine interaction which is due to
perturbative gluon exchange is
\begin{equation}\label{eq:H_HF}
H_{HF}=\frac{32 \pi}{9} \frac{\alpha_s}{m_q^2}{\bf S_1 \cdot S_2}
 \delta^3({\bf r}).
\end{equation}
Evaluating this term perturbatively with nonrelativistic wave functions
gives
\begin{equation} \label{eq:M_HF}
\Delta M_{HF}=\frac{32 \pi}{9}\frac{\alpha_s}{m_q^2} |\Psi(0)|^2
\end{equation}
for the splitting between the $\psi$ and the $\eta_c$.
The spin-spin interaction in Equation~\ref{eq:H_HF} arises from the
exchange of transverse gluons between the heavy quarks.
For massive quarks, the dominant effect of the $O(a)$ correction for
Wilson fermions, Equation~\ref{eq:WOSH}, is just such a gluon-spin
interaction, so the hyperfine splitting will be sensitive to this
correction.
According to Equation~\ref{eq:M_HF}, $\Delta M_{HF}$ is also sensitive
to the value of the quark mass, which is not determined on the lattice
to perfect accuracy.
We can therefore expect the hyperfine splitting to be a sensitive test
of lattice methods.

On the other hand, the spin averaged splitting between the lowest
angular momentum ($l=0$ and $l=1$) levels of the $\psi$ and $\Upsilon$
systems is a crucial one for lattice QCD because nonrelativistic
arguments tell us to expect it to be insensitive to these important
sources of error.
Since it is a spin averaged quantity, it should be insensitive to
uncertainties in the coefficient of the $O(a)$ correction term.
Since it is virtually the same for the $\psi$ and the $\Upsilon$,
it should be insensitive to any imperfections in our knowledge of the
quark mass.
It is therefore a good quantity to use to extract information about
QCD from lattice methods.
It may be the most accurate determination of the lattice spacing
in physical units (a key component of the extraction of the strong
coupling constant using lattice methods).

\begin{figure}
\epsfxsize=\textwidth \epsfbox{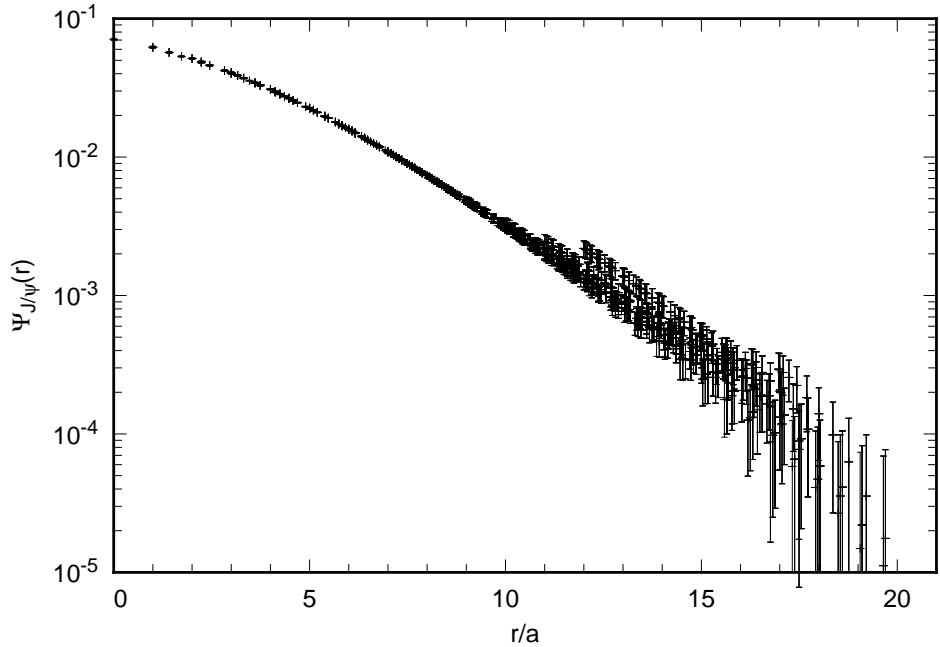}
\caption[fig:wf]{The wave function of the $J/\psi$ meson calculated on
the lattice \cite{El-Khadra+al}.}\label{fig:wf}
\end{figure}

Because the systems are nonrelativistic, their Coulomb-gauge
wave functions calculated on the lattice will give a good
picture of the properties of the states.
Figure~\ref{fig:wf} shows the wave function of the $J/\psi$ meson
calculated on a $24^4$ lattice at $\beta=6.1$ \cite{El-Khadra+al}.
It has approximately the exponential shape of a Coulomb wave function,
but at large distances it falls off faster due to confinement,
and at short distances it rises more slowly due to asymptotic freedom.
Halfway across the lattice, at $r/a=12$, the effects of periodic
boundary conditions are clearly seen.
Such wave functions have practical roles to play in lattice
calculations.
They can be used to estimate finite lattice spacing and finite volume
errors perturbatively.
They can be used to make improved operators to create and destroy the
meson states.
One of their most important roles, however, is the
clear and simple demonstration that the lattice calculations
are indeed producing charmonium states.

\begin{figure}
\epsfxsize=\textwidth \epsfbox{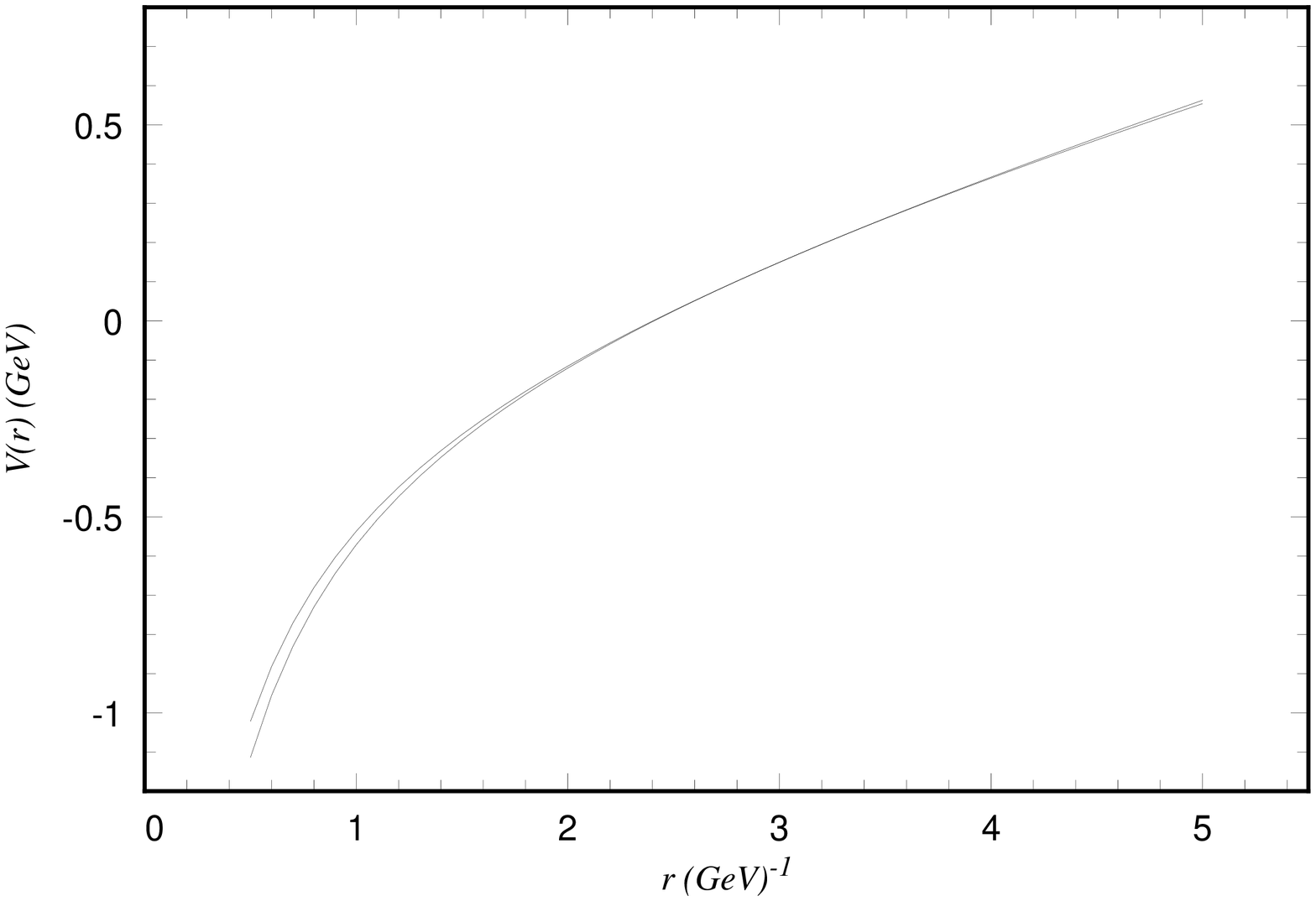}
\caption{
Results from fits to the charmonium spectrum with
asymptotically free phenomenological potentials
having the correct $\beta$ function (bottom) and the $\beta$ function
of quenched QCD (top).
}\label{fig:richpot}
\end{figure}

% \paragraph{The quenched approximation and charmonium.}

To the extent that the $\psi$ and $\Upsilon$ systems are
nonrelativistic, one can use potential model arguments to estimate and
correct for the effects of sea quarks.
These effects are expected to be rather small, since, for example, the
widths of excited $\psi$ and $\Upsilon$ states into $D$ or $B$ mesons,
50--100 MeV,  are only 10--20\% of typical energy splittings between
states.

If middle distance physics like the 1P--1S splitting is used to tune
the bare parameters of the theory,
the effective action at those distances will be about right.
In a theory with too much asymptotic freedom,
the effective coupling at short distances will be a bit too small.
Likewise, short distance quantities which depend on it, like the
wave function at the origin, will be too small.

These effects may be estimated using the Richardson potential
\cite{Richardsonpot},  which
incorporates the effects of asymptotic freedom at short distances.
Figure~\ref{fig:richpot} shows two potentials resulting from fits to
the charmonium spectrum:  the lower potential having the correct one
loop $\beta$ function $b_0 = 11 - 2n_f/3$ with $n_f=3$, and the upper
one with the stronger $\beta$ function for $n_f=0$ of quenched QCD.
As expected, the two potentials agree almost perfectly in the middle
distance region, but the $n_f=0$ potential is too soft at short
distances.

% \paragraph{The 1P--1S splitting}
The spin-averaged 1P--1S splitting has been calculated by several
groups \cite{El-Khadra+al,Lep91,NRQCDUKQCD}.
The lattice spacing in physical units is obtained by the lattice
result obtained in lattice units with the physical answer.
In the $\psi$ system, for example,
$\Delta m_{\mbox{\scriptsize 1P--1S}}=
m_{h_c}-(3m_{J/\psi}+m_{\eta_c})/4=458.6\pm0.4$~MeV.
In the $\Upsilon$ system, since the $^1P_1$ is undiscovered, the
splitting between the spin-averaged $\chi_b$ states and the
spin-averaged 1S states may be used,
$\Delta m_{\mbox{\scriptsize 1P--1S}}=452$~MeV.
The values of the lattice spacing obtained from this splitting are
shown in  Table~\ref{table:a}.
They will be crucial components in the determination of the strong
coupling constant in Section~\ref{alpha}.
They do not differ dramatically from
those obtained from other quantities, such as the $\rho$ mass
\cite{Wei92} or the string tension \cite{Bali2}.  It is the
possibility of making better uncertainty estimates that makes this
an important way of determining the lattice spacing.

In Reference~\cite{El-Khadra+al}, finite lattice spacing errors were
treated by the explicit inclusion of the term in Equation~\ref{eq:WOSH},
in the numerical calculations.
In Reference~\cite{Lep91}, corrections operators were evaluated
perturbatively using the Rich\-ard\-son potential model wave functions,
as the hyperfine splitting was evaluated in Equation~\ref{eq:M_HF}.
The lattice action of nonrelativistic QCD (NRQCD)
\cite{ThackerLepage} was used in this work,
so $O(v^2)$ correction were also included.
%The results for the $\Upsilon$ at $\beta=6.0$ are shown in
%Table~\ref{table:NRQCD}.
%The particular numbers shown in the table are not very important,
%although it is nice to see that they are not too big.
%The important fact is that potential models can be used in these
%systems to estimate systematically all of the most important sources of
%uncertainty.
These estimates could also be made without recourse to potential models,
but still using using nonrelativistic reasoning,
by using wave functions calculated directly with lattice methods
(see Figure~\ref{fig:wf}).

Both groups checked these corrections by verifying that the same answer
was obtained for several different lattice spacings.
It would be desirable to have a calculation in which both methods were
used in the same analysis in order to test carefully the method of
directly including the correction operators in the simulation, since
that is the only avenue for evaluating them available for the light
hadrons.

Likewise, for the light hadrons the functional form of the finite
volume errors in the crucial intermediate distance region
is not known, and must be calculated numerically.
Such a  calculation in a nonrelativistic system supplemented by a
nonrelativistic wave function calculation of the finite volume errors
would be useful in illuminating the methods of error analysis for the
light hadrons.

\begin{table}[t]
\caption[table:a]{Inverse lattice spacings obtained from the 1P--1S
splitting in the $\psi$ and $\Upsilon$ systems.
}
\label{table:a}
\begin{center}
\begin{tabular}{clcc}
\hline \hline
   $\beta$  & $a^{-1}$ (GeV) & System & Ref.\  	\\
\hline
5.7	&1.15(8)	&$\psi$    	& \cite{El-Khadra+al}	\\
5.9	&1.78(9)	&$\psi$	        & \cite{El-Khadra+al}	\\
6.1	&2.43(15)	&$\psi$	        & \cite{El-Khadra+al}	\\
5.7	&1.14(4)	&$\psi$	        & \cite{Lep91}	\\
5.7	&1.26(14)	&$\Upsilon$	& \cite{Lep91}	\\
6.0	&2.11(7)	&$\Upsilon$	& \cite{Lep91}	\\
\hline \hline
\end{tabular} \end{center}
\end{table}

% \paragraph{The hyperfine splitting.}

The nonrelativistic picture tells us that the hyperfine splitting and
leptonic decay amplitude are short distance quantities, proportional to
the square of the wave function at the origin.
The hyperfine splitting is also proportional to the short distance
coupling constant, and thus additionally suppressed.
The size of these suppressions for the hyperfine splitting has been
estimated as $-$30--40\% using the Richardson potential \cite{AXE2}.

The hyperfine splitting can be used to check the effects of the
$O(a)$ correction term for Wilson fermions, Equation~\ref{eq:WOSH},
which yields dominantly a spin-spin coupling for quarkonia.
Compared with its physical value of 117 MeV and the estimate of the
quenched corrected value of 70 MeV, unimproved Wilson fermions
($c=0$ in Equation~\ref{eq:WOSH}) produce splittings of as little as
10--20 MeV, depending on the lattice spacing.
Calculations with the tree level coefficient $c=1.0$
yield around 50 MeV \cite{Allton},
and with a perturbatively corrected coefficient $c=1.4 $
yield around 90 MeV \cite{AXE2}.
The precision is insufficient to allow a phenomenological determination
of the coefficient to supplant the perturbative one,
but does show clearly that the improved  action yields reasonable
results while the unimproved action does not.

% \paragraph{The Static Potential}

\begin{figure}
\epsfxsize=0.75\textwidth
\epsfbox{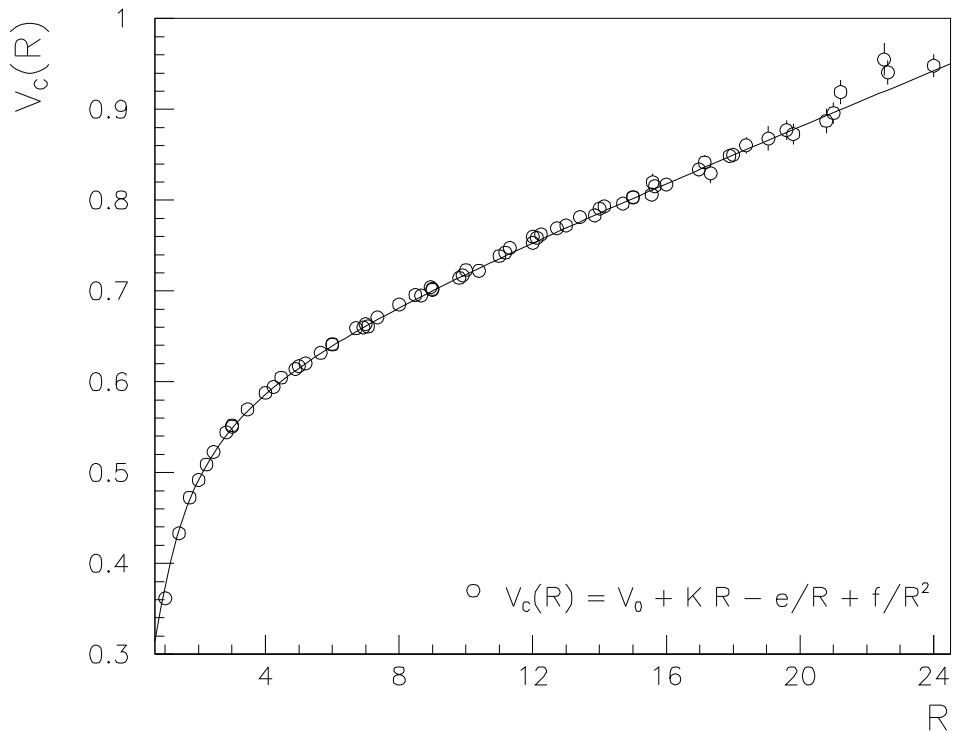}
\vspace*{3.5em}
\caption[fig:potential]{The heavy-quark potential (in lattice units)
calculated in the quenched approximation
\cite{Bali2,Bali1}.}\label{fig:potential}
\end{figure}

A topic related to quarkonium is the lattice QCD calculation of the
static potential.
In Figure~\ref{fig:potential} we show results in the quenched
approximation from a $32^4$ lattice at $\beta=6.4$ \cite{Bali2,Bali1}.
It is fit very well by a Coulomb-plus-linear form.
% as used in the original potential model work.
At short distances it agrees well with the predictions of lattice
perturbation theory \cite{lat92}.
The potential at large distances is well fit by a straight line.
The string tension obtained obeys  asymptotic scaling   to about 20\%
if a renormalized coupling constant is used.
That is, the ratio $ \sqrt{\sigma}/\Lambda^{(0)}_\MSbar $
varies by less than 20\% when $\beta > 5.7$ ($a<0.2$ fm).
Earlier apparent evidence that scaling violations as large as a factor
of two and that much smaller lattice spacings were required
has been understood as an artifact of the use of the bare lattice
coupling constant for the perturbative analysis \cite{LM2}.
This introduced poor behavior into the perturbation theory somewhat
analogous to that resulting from  attempting to do perturbative QCD
phenomenology with the MS coupling constant $\alpha_{\rm MS}(q)$ rather
than the $\MSbar$ coupling constant $\alpha_\MSbar(q)$.
The remaining small scaling violations arise from both logarithmic
(in $a$) perturbative corrections and power law finite $a$ effects,
so uncertainties associated with them cannot be removed cleanly by
extrapolation.
A summary of various recent analyses \cite{Ukawa}
contains results all falling in the range
\begin{eqnarray}
\sqrt{\sigma}/\Lambda^{(0)}_\MSbar = 1.85 \pm 10\%.
\end{eqnarray}

\subsection{The Light Hadron Spectrum} \label{spectrum}
This year Weingarten and collaborators took an important step forward
in the calculation of the light hadron spectrum in the quenched
approximation \cite{Wei92}.
This work, in a single, systematic calculation, attempted to analyze and
extrapolate away the three major source of systematic error in the
quenched approximation:  extrapolation to zero lattice spacing, to
infinite volume, and to physical quark mass.

\begin{figure}
\epsfxsize=\textwidth \epsfbox{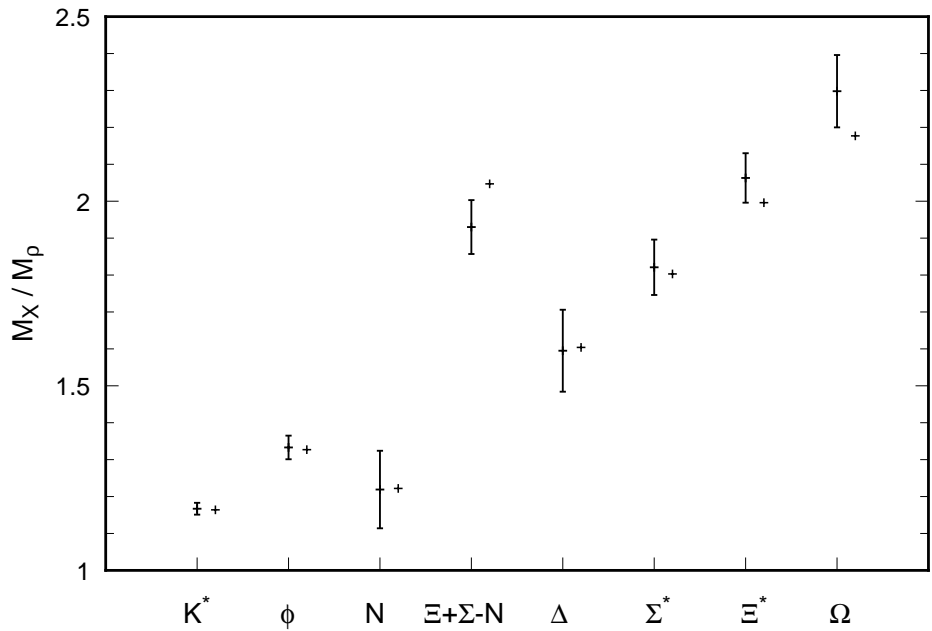}
\caption[fig:spectrum]{The spectrum of the light hadrons.
Error bars are from lattice calculations in the quenched approximation
\cite{Wei92}, and $+$ denotes experiment.}\label{fig:spectrum}
\end{figure}

The results are shown in Figure~\ref{fig:spectrum}.
The lattice spacing has been eliminated using $m_\rho$, and the bare
quark masses using $m_\pi^2$ and $m_K^2$.
The errors shown are statistical.
The authors argue that the uncertainties involved in the extrapolations
to infinite volume and physical quark mass are smaller than the
statistical errors.
They have not attempted to estimate the uncertainty in the extrapolation
to zero lattice spacing.

The extrapolation of $m_{\pi}^2$ and the masses of the other hadrons
to physical quark mass was made linearly in $m_q$
 in accordance with theoretical prejudice.
The expected functional forms were seen in the data.
A version of the Gell-Mann--Okubo formula was used to argue that the
error arising from this extrapolation was around 1\%.
This estimate could be supplemented by direct numerical investigation
of the functional form.

The extrapolation to infinite volume was made by performing the
calculations on the coarsest lattice at
several volumes, and using the results to extrapolate the calculations
on the finer lattices to infinite volume.
The extrapolation used only two points.
Much work is now by several groups \cite{Ukawa}
 to determine the functional form
of the volume dependence which should make the extrapolation reasonably
solid.

The results were extrapolated linearly in $a$ to zero lattice  spacing
in accordance with theoretical prejudice, but the data were not precise enough
to test the accuracy of that prejudice.
This extrapolation can be improved by adding the
 single additional term to the quark action which suffices to remove the
$O(a)$ error from Wilson fermions \cite{WOSH}
and checking that the observed dependence of the results on the lattice spacing
disappears.

Possible problems with contamination from
higher mass states in each had\-ron\-ic channel
showing up in the work of other groups have been emphasized
by Ukawa \cite{Ukawa}.
An important contribution toward reducing these problems was made by the
APE collaboration in 1988 \cite{APE88} who pointed out that quarks
spread out over roughly the size of a light hadron have a much larger
overlap with the light hadrons and a much smaller overlap with excited
states than do the local quark operators which had been in use up to that
time.
Much more sophisticated work along these lines is possible following the
lead of glueball calculations whose worse signal to noise problems have
forced a more serious examination of this problem \cite{Kronfeldwf}.

The analysis of the uncertainties in a light hadron calculation is more
demanding than in a charmonium calculation,
since nonrelativistic arguments do not help.
Nevertheless, there are further technical tools available,
such as Equation~\ref{eq:WOSH},
than have
been applied so far in this calculation, which should make possible
the confirmation or improvement of these uncertainty estimates
for the quenched light hadron spectrum
with the present generation of computers.

\section{QCD Phenomenology}\label{QCD}
QCD is supposed to describe high-energy perturbative phenomena, such as
deep inelastic scattering, as well as low-energy nonperturbative
phenomena, such as hadron masses.
In QCD with $n_f$ quark flavors, there are $n_f+1$ parameters, the quark
masses and the strong coupling constant.
The latter is equivalent to a standard mass to set the scale in MeV.
Using the nonperturbative lattice formulation of QCD it is possible to
compute them using $n_f+1$ hadron masses as the physics input.
If these masses are chosen unwisely, a cumbersome juggling act must be
performed, adjusting the bare parameters, computing the hadron masses
and working back to the renormalized parameters.
As explained in Section~\ref{onia} the 1P--1S splitting of quarkonium is
insensitive to light and heavy quark masses.
Consequently, it is ideal for converting from lattice units to MeV.
% By application of the renormalization group, setting the scale in this
% way gives physical meaning to $\alpha_S(q)$.
Once this has been done, it is relatively straightforward to use meson
masses to determine quark masses.

Given $\alpha_S$ one can then test whether the same QCD describes the
strong interactions at all energies.
One simply inserts the nonperturbatively computed coupling into the
perturbative series for high-energy scattering and compares with data.
A favorable outcome will increase our quantitative confidence
in QCD enormously.
At present lattice QCD can offer $\alpha_S$ with systematic uncertainties
comparable to deep inelastic scattering, although the analysis is
less mature.
These results, and the theoretical ideas needed to reduce the
uncertainties to a negligible level, are in Section~\ref{alpha}.

Because of confinement, quark masses cannot be measured directly.
However, every serious theoretical construct that goes beyond the
standard model provides $\MSbar$ quark masses as an output.
Hence, good estimates of quark masses from lattice QCD should prove
useful to builders of new physics models.

\subsection{The Coupling Constant}\label{alpha}

The numerical value of the strong coupling depends on the ``scheme''
chosen to define it.
A scheme can be defined by a renormalization convention, such as the
$\MSbar$ scheme in dimensional regularization or the bare scheme in
lattice perturbation theory.
More generally, it can be defined by any physical quantity that is equal
to the bare coupling at the leading order of perturbation theory.
For example, in QED the low-energy limit of Thomson scattering is used
to define the electromagnetic coupling.

%A determination of the strong coupling constant with lattice methods
%consists of two elements:

%and the calculation of some dimensionful quantity such as
%$\Delta m_{\mbox{\scriptsize 1P--1S}}$ to fix the lattice scale in
%physical units.

%The static potential is not

For QCD L\"uscher, et al., have delineated four criteria
for a practical scheme \cite{Lue93}.
The physical quantity should have a rigorous nonperturbative
definition; otherwise it cannot be calculated nonperturbatively.
In Monte Carlo simulations it ought to have a good signal-to-noise
ratio, so that small statistical uncertainties can be achieved in a
reasonable amount of computer time.
Furthermore, uncertainties arising from extrapolations in lattice
spacing and quark mass should accumulate slowly.
Finally, a perturbative calculation of the physical quantity must be
tractable, so that the nonperturbative coupling can be used in
perturbative QCD.

Once one has chosen a scheme $s$, one must relate the dimensionless
coupling to the standard mass.
This is done using the renormalization group.
It is important to realize that renormalization-group calculations can
be carried out nonperturbatively.
Because of asymptotic freedom, the nonperturbative and perturbative
$q$ dependence must agree for large enough $q$.
The region of agreement provides a numerical value for $g_s^2(q)$ that
can be used in perturbative series for high-energy scattering.
The standard mass is needed to convert $q$ from lattice units of
the nonperturbative calculations to physical units.
For example,
\begin{equation}
q~({\rm MeV})=\frac{aq}{a\Delta m_{\mbox{\scriptsize 1P--1S}}}
458.6~({\rm MeV});
\end{equation}
the numerator and denominator of the fraction come from the lattice
calculation, and
$\Delta m_{\mbox{\scriptsize 1P--1S}}=458.6~{\rm MeV}$
from experiment. %  these guys did $h_v$ only \cite{Arm92}.

%Early Monte Carlo renormalization groups imagined taking the scale $q$
%near the cutoff \cite{Wil79}.
%Then an obvious choice of the scheme is the bare coupling.
%For the standard action this turns out to be a poor choice.
%A better choice is to relate $g^2_\MSbar$ to a short-distance quantity
%that can be calculated numerically to good precision \cite{LM2}.

In Reference \cite{El-Khadra+al}, a perturbative relation
(improved by mean field theory) was used to estimate
the continuum coupling constant in terms of the bare lattice coupling constant.
In Reference \cite{LM2} it was found that perturbative calculations
of short distance physical quantities in terms of a coupling constant
estimated in this way were systematically lower
than Monte Carlo calculations by a few per cent.
This suggests that a slightly improved determination would be given by
extracting the coupling directly from Monte Carlo calculations.
It remains to be checked that there are no substantial deviations
between the couplings determined in this way, from physics on scales
from one to half a dozen lattice spacings,
 and true continuum
couplings \cite{Lue93}.

The work of Reference \cite{El-Khadra+al} gave
\begin{equation}\label{alpha-MSbar-0}
\alpha_\MSbar^{(0)}(5~{\rm GeV})=0.140\pm0.004,
\end{equation}
where the superscript emphasizes the number of active quark flavors.
This error bar comes from the statistical uncertainty in the 1P--1S
splitting used to determine $a$, augmented somewhat by lattice-spacing
effects.
Complementary analyses based on the short-distance static potential
\cite{Bali2,Mic92} yield values of $\alpha_\MSbar^{(0)}(5~{\rm GeV})$
consistent with these.

The $n_f=0$ result can be converted to $n_f=4$ by appealing to the
potential models that describe quarkonia so well.
Choosing the 1P--1S splitting to set the scale is equivalent to
adjusting the coupling of the quenched theory to reproduce physics at
intermediate energies.
Since the coupling runs faster for fewer quarks, this adjustment makes
the coupling at 5~GeV too small.
Correcting the quenched result for this effect yields \cite{El-Khadra+al}
\begin{equation}\label{alpha-MSbar}
\alpha_\MSbar^{(4)}(5~{\rm GeV})=0.174\pm0.012.
\end{equation}
For comparison with the compilation in Reference~\cite{PDG92}
\begin{equation}\label{alpha-MSbar-Z}
\alpha_\MSbar^{(5)}(M_Z)=0.105\pm0.004.
\end{equation}
The error bar in Equation~\ref{alpha-MSbar} is three times larger than
in Equation~\ref{alpha-MSbar-0}, because the matching energy is not
known exactly, and because for charmonium it is rather low.
The bulk of the correction is due to the effects of light quarks
on the potential at short distances, which can be calculated in
perturbation theory.
However, a part of the correction arises from the effects of light
quarks on the potential at middle distances, which must be estimated
phenomenologically.

It is therefore significant that
a similar analysis has also been carried out using NRQCD in both the
$\psi$ and $\Upsilon$ systems \cite{Lep91}.
Typical energy scales in the $\Upsilon$ are about twice those in the
$\psi$.
(For example, typical gluon momenta are 400~MeV and 800~MeV in the
$J/\psi$ and $\Upsilon$ states, respectively.)
The effects of light quarks in the murky intermediate distance region
may be expected to be quite different at the
$\Upsilon$ than at the $\psi$.
Although some details of the systematic error analysis is different,
the $\psi$-system calculation agrees with Equations~\ref{alpha-MSbar-0}
and \ref{alpha-MSbar}.
There are subtle differences in the determination of $\alpha_\MSbar$
from the $\Upsilon$ system, which arise because the typical energy
scales are higher.
As shown in Table~\ref{table:a} the lattice spacing determined by the
$\Upsilon$ 1P--1S splitting is about 10\% smaller.
Propagating this change implies that $\alpha_\MSbar^{(0)}(5~{\rm GeV})$
is somewhat larger.
However, the correction for the quenched approximation is smaller,
because the matching is done at somewhat higher energies.
If the argument used to compute the correction is valid, the two
effects should cancel in $\alpha_\MSbar^{(4)}$.
Reference~\cite{Lep91} finds
\begin{equation}\label{alpha-MSbar-Upsilon}
\alpha_\MSbar^{(4)}(5~{\rm GeV})=0.170\pm0.010
\end{equation}
from the $\Upsilon$ system, which agrees remarkably well with the
results from the $\psi$ system.

The only way to eliminate the error from the quenched approximation in
Equation~\ref{alpha-MSbar} is to perform calculations in full QCD.
The second-most important uncertainty comes from the dependence on the
lattice spacing.
Because the scale $q$ is tied to the cutoff in these calculations,
it is impossible to separate the scaling dependence from any other $a$
dependence.
In other words, the criterion that uncertainties do not accumulate
during extrapolation is not strictly respected.
To clear things up, one must associate $q$ with a physical scale.
An elegant way to do so is to take $q=1/L$ \cite{Lue92},
where $L$ is the linear size of the finite volume.
References~\cite{Lue93,Lue92} also suggest a class of schemes for which the
extrapolation to the continuum limit is controlled.
So far these ideas have been applied to the pure SU(2) gauge theory
\cite{Lue93,Lue92}.
For the coupling chosen, the scaling behavior matches two-loop
perturbation theory at surprisingly low energies, perhaps even as low as
$q=1$~GeV.

In several years full QCD calculations with the scaling analysis of
References~\cite{Lue93,Lue92} will have computed the strong coupling constant
with a precision of a few per cent.
The uncertainty will be due to finite statistics, compounded somewhat by
extrapolations to zero lattice spacing and physical quark masses.
There will be no uncertainty from truncating perturbation theory and no
uncertainty from nonperturbative effects.
The specific value for $\alpha_s(q)$ will be complemented by an energy
scale $q$, above which perturbative evolution is valid.
Purely perturbative calculations can then be used to relate the
nonperturbative scheme $s$ to the $\MSbar$ scheme.
Rather than use this relation to {\em determine\/} $\alpha_\MSbar$, one
ought to {\em eliminate\/} it from high-energy perturbative series
in favor of $\alpha_s$.
This is analogous to the strategy used in perturbative QED, where the
$\MSbar$ coupling is used only as an intermediate step.

\subsection{Quark Masses}\label{quark-mass}
The masses of the charm and bottom quarks are currently estimated from
potential model calculations.
Lattice calculations should eventually be able to pin these down
to a precision of perhaps 5\%, limited by perturbation theory.
The existing numerical data on the masses of the $J/\psi$ and the
$\Upsilon$ is already quite adequate for this purpose.
The remaining work required is short-distance lattice perturbation
theory with massive quarks, which is rather complicated.

The top quark is expected to decay weakly before it can form a QCD
bound state.  Lattice calculations are unlikely to be useful
in determining its mass after it is found.

For the light quarks it is convenient to discuss the combinations
$\hat{m}=\half(m_d+m_u)$, $\Delta m^2_{du}=m^2_d-m^2_u$, and $m_s$.
Ratios of the light-quark masses are currently best estimated using
chiral perturbation theory, a systematic description of the low energy,
small quark mass limit of QCD \cite{Gas82}.
To set the overall scale requires a dynamical calculation.
In lattice QCD, $\hat{m}$ and $m_s$ can be extracted from the variation
in the square of the pseudoscalar mass between $m_\pi^2$ and $m_K^2$.
The most difficult quark-mass combination is $\Delta m^2_{du}$, which
causes the isospin-violating part of the splittings in hadron multiplets.
Since chiral perturbation theory provides a formula for
$\Delta m^2_{du}/m_s^2$ with only second-order corrections, it is
likely that the best determination of $\Delta m^2_{du}$ will come from
combining the formula with a lattice QCD result for $m_s$.

\begin{figure}
\epsfxsize=\textwidth \epsfbox{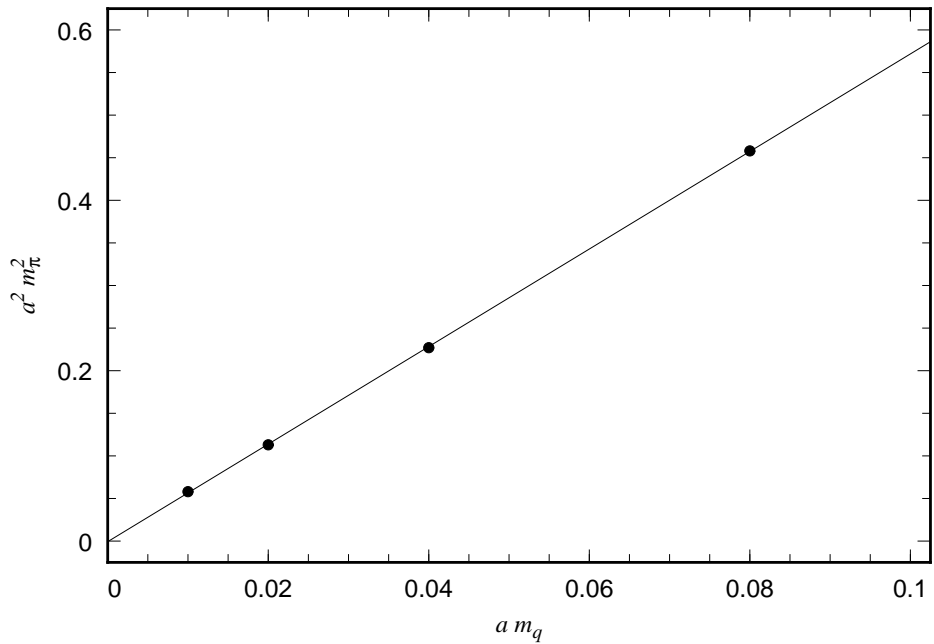}
\caption[fig:chiral]{Pseudoscalar meson mass squared as a function of
the quark mass, calculated in the quenched approximation \cite{Cab91}.
The line is a linear fit to the data.}\label{fig:chiral}
\end{figure}

Uncertainties in the chiral estimates of these ratios arise from
varying treatments of higher-order terms.
Existing lattice calculations either use very massive sea quarks or
ignore sea quarks entirely, so they also treat higher-order hadronic
effects somewhat incorrectly.
We probably must wait for better calculations including sea quarks
correctly before lattice calculations can contribute to the
determination of the ratios.
As illustrated in Figure~\ref{fig:chiral}, present calculations show a
linear relation between the quark mass and the square of the meson mass,
as expected from lowest order chiral perturbation theory alone, up to
surprisingly large values of the quark mass.

The overall mass scale of the light quarks is currently
determined by less reliable phenomenological assumptions.
A full error analysis of lattice determinations of this quantity has
not been completed, but it is poorly enough known from conventional
phenomenology that it is worth discussing the state of the lattice
results.
Because staggered fermions have an exact chiral symmetry, they
are likely to be superior for these calculations.
As summarized by Ukawa \cite{Ukawa}, quenched results for
$\hat{m}_\MSbar(1~{\rm GeV})$ are in the range 2.4--3.0~MeV.
This is outside the range of 3.5--11.5~MeV indicated in
Table~\ref{table:sm}.
There is no evidence in the data for large finite volume, finite
lattice spacing, or statistical errors.
When relatively heavy sea quarks are added to the calculation no qualitative
change is observed: these results cluster around 2~MeV.
These results should not be taken too seriously until a more complete
error analysis exists, but the possibility that the conventional
estimates are too high is intriguing.

\section{Weak-Interaction Phenomenology}\label{WIP}
We now turn to the role lattice QCD can play in determining the
Cabibbo-Kobayashi-Maskawa (CKM) matrix.
In the standard model the CKM matrix accounts for four of the 19
parameters.
Furthermore, to test the standard model one would like to extract all
elements of the CKM matrix and verify that it is unitary.
Because the observable consequences of the CKM matrix involve weak
transitions of hadrons, nonperturbative QCD enters immediately.
We shall focus on processes that are especially amenable to lattice
technology that also play a crucial role in determining the CKM matrix.
For a review of weak-interaction phenomenology with emphasis on the CKM
matrix see Reference~\cite{Nir92}; for more technical reviews of the lattice
technology see References~\cite{Ber92,Sachrajda92}.
This section discusses leptonic decays in subsection~\ref{decay-constant},
semi-leptonic decays in subsection~\ref{semi-leptonic}, and neutral meson
mixing in subsection~\ref{mixing}.
A brief explanation of the difficulties with non-leptonic decays is in
subsection~\ref{non-leptonic}.

A lattice large enough to encompass the scales $\Lambda_{\rm QCD}$ and
$m_W$ (or $m_t$) would have several thousand sites on each side.
That is obviously not feasible.
Fortunately, it is also not necessary.
Leptonic and semi-leptonic decay amplitudes factor into a product of
leptonic and hadronic matrix elements of electroweak currents.
Lattice QCD is needed to calculate the hadronic factors
$\langle 0 |J_\mu|h\rangle$ for leptonic decays and
$\langle h'|J_\mu|h\rangle$ for semi-leptonic decays.
For neutral meson mixing and non-leptonic decays, the standard
theoretical apparatus uses the operator product expansion to disentangle
contributions above and below a scale $\mu<m_W~({\rm and}~m_t)$.
This analysis leads to the effective weak Hamiltonian, which can be
written schematically as
\begin{equation}\label{weak-OPE}
H_{\rm eff} = \sum_n C_n(\mu) O^{(n)}(\mu),
\end{equation}
where, to leading order in $m_W^{-2}$, the $O^{(n)}$ are four-quark
operators.
Lattice QCD is needed to calculate hadronic matrix elements
$\langle h'|O^{(n)}|h\rangle(\mu)$, where $h$ and $h'$ are various
hadronic states.

Several conditions must be met before hadronic matrix elements of the
four-quark operators can be applied to phenomenology.
The $\mu$ dependence of the coefficient functions and the four-quark
operators must cancel.
Since the coefficient functions are determined perturbatively, the
lattice calculations must be performed with lattice spacings for which
perturbation theory is applicable.
In this way the lattice regulated matrix element
$\langle h'|O^{(n)}_{\rm lat}|h\rangle(\pi/a)$
can be related to the renormalized matrix element
$\langle h'|O^{(n)}_R|h\rangle(\mu)$
in the scheme $R$ and at the scale $\mu$ for which the coefficient
functions are available.
Similar, but simpler, relations apply to currents $J_\mu$ as well, see
below.
With our present understanding of lattice perturbation theory
\cite{LM2}, this conversion should not introduce large uncertainties.

Independent of such scheme and scale dependence, the four-quark
operators must be defined nonperturbatively.
Interactions cause mixing with operators with the same (lattice)
quantum numbers.
When these other operators have the same or higher dimension, it is
presumably adequate to use the definitions of perturbation theory.
It is at least consistent, because a similar classification already
arises in the operator product expansion, Equation~\ref{weak-OPE} which
is established perturbatively.
A more pernicious problem is mixing with {\em lower\/} dimension
operators.
Since the coefficients of these operators contain inverse powers of $a$
there is no reliable method to remove them perturbatively.
The only feasible way to remove them nonperturbatively is to insist on
the correct scaling behavior and the restoration of continuum-limit
symmetries.

\subsection{Decay Constants}\label{decay-constant}
Lattice calculations of the decay constants are necessary both as tests
and as predictions of lattice QCD.
We shall follow the normalization convention that for pseudoscalar
mesons
\begin{equation}\label{f_pi}
\langle0|\bar{u}\gamma_\mu\gamma_5 d|\pi^-\rangle=ip_\mu f_\pi
\end{equation}
and for vector mesons
\begin{equation}\label{f_rho}
\langle0|\bar{u}\gamma_\mu d|\rho^-;\lambda\rangle=
i\varepsilon_\mu^{(\lambda)}m_\rho f_\rho.
\end{equation}
{}From leptonic decays one finds $f_\pi=131$~MeV, $f_K=160$~MeV, and
$f_\rho=216$~MeV.
On the other hand, the decay constants of heavy-light mesons ($D$ and
$B$) are not known experimentally, and the measurements would be
difficult.
Hence, even semi-quantitative lattice calculations of $f_D$ and $f_B$
are interesting, and, when the uncertainties are fully understood,
quantitative lattice calculations will play an essential role in
understanding $D$- and $B$-meson phenomenology \cite{Eic87}.

Section~\ref{methods} pointed out that the lattice artifacts of masses
approach the continuum limit as a power of $a$.
Hadronic matrix elements, such as decay constants, typically approach
the continuum limit more slowly.
The lattice operator and the continuum operator are related as follows
\begin{equation}\label{lat-cont}
\left.\bar{u}\gamma_\mu\gamma_5 d\right|_{\rm lat}=
\left.\bar{u}\gamma_\mu\gamma_5 d\right|_{\rm cont}+
c' a \bar{u}D_\mu\gamma_5 d+\cdots,
\end{equation}
where $a$ is the lattice spacing.
In one-loop perturbation theory one easily sees that $D_\mu$ in the
second term can absorb a gluon with momentum $\sim1/a$.
This contribution, together with analogous ones from the unwritten
terms, yield $c g_0^2 a (1/a)/(16\pi^2)$.
Generalizing to all orders
\begin{equation}\label{lat-cont-mx}
\left.\bar{u}\gamma_\mu\gamma_5 d\right|_{\rm lat}=
\left.\bar{u}\gamma_\mu\gamma_5 d\right|_{\rm cont} \left[1 +
\sum_{\nu=1}^\infty c_\nu\left(\frac{g_0^2}{16\pi^2}\right)^\nu
\right] + {\rm O}(a),
\end{equation}
where equality holds for matrix elements of low-momentum states,
and ${\rm O}(a)$ denotes terms that vanish as a power.
For small $a$, $g_0^2\propto(\log a)^{-1}$.
Hence $\bar{u}\gamma_\mu\gamma_5 d|_{\rm lat}$ approaches
$\bar{u}\gamma_\mu\gamma_5 d|_{\rm cont}$ rather slowly.
Although we have used the axial current as an example, composite
operators generally obey an equation analogous to
Equation~\ref{lat-cont-mx}.

There are several strategies for handling the lattice-spacing errors
indicated in Equation~\ref{lat-cont-mx}.
One can ignore the perturbative bracket and hope that the ${\rm O}(a)$
terms are the largest lattice artifact at accessible values of $a$.
This would only be sensible if the coefficients $c_1$ were small, but
explicit calculations in several papers \cite{Mey83,Mar83,Gro84} show
that they are not.
One could acquire numerical data over a wide range of $g_0^2$ to
perform a correct extrapolation, but that is impractical.
Fortunately, it is possible to improve the situation.
First, if one recasts perturbative series such as the one in
Equation~\ref{lat-cont-mx} in terms of a renormalized coupling constant,
one expects the higher-order corrections to be small \cite{LM2}.
Second, most of $c_1$ comes from a certain class of diagrams (Feynman
gauge tadpole diagrams) \cite{Kro85}.
These contributions can be isolated and treated nonperturbatively
\cite{LM2}.
Third, systematic improvement to the action \cite{WOSH} and the
operators \cite{Hea91} can reduce the ${\rm O}(a)$ terms.
With these three improvements it should be possible to reduce
lattice-spacing errors so that they are smaller than the statistical
uncertainties.

The most systematic investigation of light-meson decay constants
\cite{Wei93} uses the same gauge configurations and quark propagators
used to compute light-hadron masses in Reference~\cite{Wei92}.
The extrapolation in quark mass and the finite-volume corrections were
handled in the same manner as for the hadron masses
(cf.\ Section~\ref{spectrum}).
In this case the finite-volume corrections increase the error bars.
The lattice-spacing extrapolation was done as follows. % two ways.
The logarithmic $a$ dependence and some of the ${\rm O}(a)$ dependence
was accounted for as specified in References~\cite{LM2,KMf_B,KMAms}, and
the remaining $a$ dependence was assumed to be linear in $a$.
The results of this analysis are tabulated in Table~\ref{decay-table}.

\begin{table}[t]
\caption[decay-table]{Summary of results for decay constants.
The error bars for light mesons \cite{Wei93} do not include
errors estimates for the quenched approximation and plausibly small
residual lattice-spacing errors.
The error bars for heavy-light mesons \cite{Lab93} do not include,
quenched, finite-volume, or non-zero lattice spacing errors.
See the text for a discussion of these errors.}
\label{decay-table}
\begin{center}
\begin{tabular}{cccccc}
\hline \hline
        & $f_\pi/m_\rho$ & $f_K/m_\rho$ & $f_\rho/m_\rho$ &
$f_D/f_\pi$ & $f_B/f_\pi$ \\
\hline
{expt.} & $0.171$ &  $0.209$  &   $0.281$  &       &       \\
 lQCD   & $0.129^{+0.040}_{-0.051}$ & $0.164^{+0.030}_{-0.034}$
        & $0.245^{+0.055}_{-0.049}$ & $1.58\pm0.15$ & $1.43\pm0.15$ \\
{Ref.}  & \cite{Wei93} & \cite{Wei93} & \cite{Wei93} &
           \cite{Lab93} & \cite{Lab93}  \\
\hline \hline
\end{tabular} \end{center}
\end{table}

One of the most eagerly pursued topics in lattice QCD is the calculation
of heavy-light meson properties.
When one of the quarks in the meson becomes heavy, the dynamics
simplifies considerably \cite{Eic79,Isg89}.
In particular, for $m_q\gg\Lambda_{\rm QCD}$ the typical momentum in a
heavy-light meson remains small, $p\sim\Lambda_{\rm QCD}$.
The energy scale $m_q$ decouples from the heavy quark dynamics, making
it possible to derive effective theories
% \cite{Cas86,Eic90,Isg90,Lep87,Geo90,Gri90,Lep92}.
\cite{Cas86}--\cite{Lep92}.
For infinite mass there are new symmetries among different spins and
flavors of heavy quarks.
These symmetries have many interesting implications.
For example, $m_P=m_V$ and $f_P=f_V$, where ``$P$''and ``$V$''
denote generic pseudoscalar and vector heavy-light mesons, and the
various form factors discussed in Section~\ref{semi-leptonic} can be
expressed in terms of one universal function \cite{Isg90}.

For this section the most important result of heavy-quark symmetry is
a scaling law for the pseudoscalar decay constant
$f_P\propto M_P^{-1/2}$ \cite{Shi87}.
The leading symmetry-breaking effect is at order $M_P^{-1}$, i.e.\
\begin{equation}
\Phi_P=f_P\sqrt{M_P} = \Phi_\infty - \Phi'_\infty M_P^{-1}.
\end{equation}
Because of the theoretical utility of heavy-quark symmetry, lattice QCD
results for $\Phi_\infty$ and $\Phi'_\infty$ are interesting, as well as
the physical results $f_D$ and $f_B$.

The large mass is also an important technical issue for lattice QCD
calculations of heavy-light meson properties.
At currently accessible values of the lattice spacing, charm and bottom
lie in region $m_qa\approx1$, and for the infinite mass limit one must
reconcile $m_q\rightarrow\infty$ and $a\rightarrow0$ in a compatible
way.
This is done by formulating a lattice action of the effective theories,
either static \cite{Eic87} or nonrelativistic \cite{Lep87,Lep92}.
In analogy with eqs.~(\ref{lat-cont}) and (\ref{lat-cont-mx}), the
currents of the effective lattice theory must be matched to the
relativistic continuum theory \cite{Hil90,BLP89,Dav92,Mor93}.
Another approach is to use Wilson fermions and extrapolate towards
infinite mass.
At first sight this seems risky.
However, it is possible to show how the energy scale $m_q$ decouples in
the lattice theory \cite{Kro93}.
Such an analysis shows how to interpret the Wilson theory as
an effective theory, and how it shares many features with the static
and nonrelativistic theories \cite{KMf_B,KMAms}.

Now let us discuss results from lattice QCD for $\Phi_\infty$,
$\Phi'_\infty$, $f_D$, and $f_B$.
Most of the work has focussed on one of two lines of attack.
One is a systematic analysis of the infinite-mass (or static) limit
\cite{EHT91,Ale93,Bou89,All91}, concentrating on $\Phi_\infty$.
The important technical issues are optimizing the signal-to-noise ratio,
and studying the lattice-spacing and finite-volume dependence of
$\Phi_\infty$.
% When these i    it will be possible to evaluate $\Phi'_\infty$ by
% treating the heavy-quark kinetic energy as a perturbation \cite{EHi90}.
The other line of attack is to concentrate on the mass dependence.
Until now this has meant combining numerical data from quark masses near
the charm mass with the static-limit results, and interpolating
\cite{Lab93,Lab91,Ale92,Aba92,Ric93}.

Results for $f_D$ and $f_B$ from Reference~\cite{Lab93} are in
Table~\ref{decay-table}.
(We cite Reference~\cite{Lab93} because it comes close to incorporating
the mass effects derived in References~\cite{KMf_B,Kro93}.)
Heavy-strange meson decay constants are
$f_{D}/f_{D_s}=f_{B}/f_{B_s}=0.90\pm0.05$.
Taking meson masses and $f_\pi$ from experiment, the scaling
combinations are
$\Phi_D=0.28\pm0.03~{\rm GeV}^{3/2}$ and
$\Phi_B=0.43\pm0.04~{\rm GeV}^{3/2}$,
which can be compared with the static limit
$\Phi_\infty=0.53\pm0.10~{\rm GeV}^{3/2}$ \cite{Lab93}.
(This value is consistent with References~\cite{Ale93,Ric93} and with
Reference~\cite{EHT91} when scale-setting ambiguities are resolved.)
The systematic studies of the static limit \cite{Ale93} suggest that the
extrapolation to infinite volume will change these results negligibly,
and that the extrapolation to $a=0$ may reduce the results by 10\%.
A more serious source of uncertainty comes from setting the scale.
The results presented here use $f_\pi$ to set the scale.
This may not be the best choice as a rule, but one might argue that the
quenched approximation's errors cancel to some extent in $f_P/f_\pi$.
For example, the ratio $f_K/f_\pi$ in Table~\ref{decay-table}
\cite{Wei93} agrees much better with experiment than the decay
constants themselves.

Although the numerical results may not yet be definitive, there are two
important conclusions to draw from these lattice results:
First, $\Phi_\infty$, $f_D$, and $f_B$ are larger than many
model calculations had suggested \cite{Ros90}.
By combining the first column of Table~\ref{decay-table} with the
heavy-light results, one sees that discrepancy is even more dramatic
using $m_\rho$ as the standard of mass.
Second,
the $1/M_P$ corrections are large and phenomenologically important.
This is not really unexpected from the heavy-quark symmetry arguments,
since the correction is first order, and it need not be indicative of
the size of second-order corrections.

\subsection{Semi-Leptonic Decays} \label{semi-leptonic}
\begin{figure} % [b]
\epsfbox{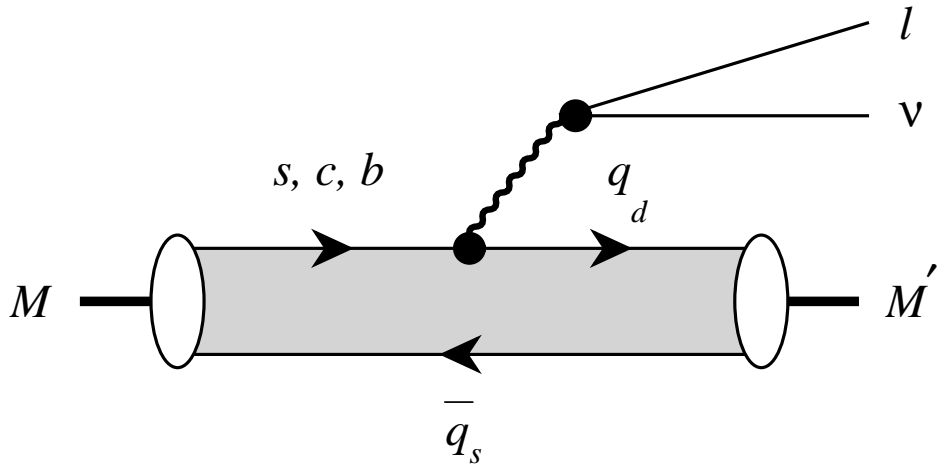}
\caption[quarkflow]{
Spectator diagram for meson semi-leptonic decays.
For the weak interactions, the diagram may be interpreted as a Feynman
diagram.
However the strong interactions binding quarks into mesons must be
treated nonperturbatively, as indicated by the grey shading.
}\label{quarkflow}
\end{figure}

A generic semi-leptonic decay can be denoted
$A\rightarrow Xl \nu$, where $A$ is a flavored hadron.
We shall focus on mesons, because they are easier than baryons to study,
both experimentally and theoretically.
The process%
\footnote{When the final state meson $X$ is an isoscalar and $A$ is
charged, there is another diagram in which $A$ annihilates into $W$ and
$X$ emerges out of the glue.
For simplicity we shall ignore these decays.}
is depicted in Figure~\ref{quarkflow}.
The flavored quark (strange, charm, or bottom) undergoes a weak decay
by emitting a virtual $W$ boson that subsequently turns into the lepton
pair.
The other quark ($\bar{q}_s$ in Figure~\ref{quarkflow}) does not take
part in the weak decay, so it is called the spectator.
However, the QCD interaction between the spectator quark and the
decaying quark ($s,c,b\rightarrow q_d$ in Figure~\ref{quarkflow}) is the
most difficult feature of the decay to calculate.
It is the part that requires lattice QCD.

The amplitude for $A\rightarrow Xl \nu$ is proportional to the
hadronic matrix element $\langle X|J_\mu|A\rangle$, where $J_\mu$ is
the $V-A$ charged current.
If the quark of flavor $a$ turns into flavor $x$
\begin{equation}
J_\mu = \bar{x}\gamma_\mu(1-\gamma_5)a.
\end{equation}
It is convenient to express the amplitude in terms of form factors.
When $X$ is a pseudoscalar meson one writes
\begin{equation}
\langle X|J_\mu|A\rangle=f_+(q^2)(p+p')_\mu + f_-(q^2)(p-p')_\mu,
\end{equation}
where $p$ ($p'$) is the initial (final) state meson's momentum and
$q=p-p'=p_l +p_\nu$.
Similarly, when $X$ is a vector meson there are four independent form
factors.
Decays correspond to the kinematic region
$m_l^2<q^2\leq q_{\rm max}^2=(m-m')^2$; in the rest frame of the
initial meson, the neutrino is soft at $q^2=m_l^2$, whereas the
final state meson is at rest at the ``endpoint'' $q^2=q_{\rm max}^2$.

The interplay between experiment, lattice QCD, and the CKM matrix
becomes clear upon examining the differential decay rate.
For example, when $X$ is a pseudoscalar meson
\begin{equation}\label{semi-leptonic-rate}
\frac{d\Gamma}{dq^2} =
\frac{G_F^2\lambda^{3/2}}{192\pi^3m_A^3}\left|V_{ax}f_+(q^2)\right|^2,
\end{equation}
where $V_{ax}$ is the element of the CKM matrix associated
with the quark-$W$ vertex in Figure~\ref{quarkflow}, and
$\lambda = (m_A^2-m_X^2 - q^2)^2 - 4m_X^2q^2$.
The contribution of $f_-$ to the rate is proportional to the lepton
mass, so in most cases it can be neglected.
The exceptions are $K\rightarrow\pi\mu\nu$ and $\tau$ lepton final
states.
When $X$ is a vector meson, the decay rate obeys a similar formula, with
the contribution of one of the four form factors suppressed by one power
of the lepton mass.
Everything in Equation~\ref{semi-leptonic-rate} is well-known or measurable
except $V_{ax}$ and $f_+$, so a measurement of $d\Gamma/dq^2$
constitutes a measurement of $|V_{ax}f_+(q^2)|$.
Specific decays and their CKM matrix elements are shown in
Table~\ref{semi-CKM}.
\begin{table}[t]
\caption[semi-CKM]{Semi-leptonic decays and the CKM matrix elements they
determine.
For brevity only pseudoscalar final states are listed; vector final
states are $\rho$, $K^*$ and $D^*$, as appropriate.}\label{semi-CKM}
\begin{center} \begin{tabular}{r@{$\,\rightarrow\,$}lcl}
\hline \hline
$A$ &  $X$  & $V_{ax}$ & \multicolumn{1}{c}{\sc comment} \\
\hline
$K$ & $\pi$ & $V_{us}$ & calibrate quenched approximation      \\
$D$ & $\pi$ & $V_{cd}$ &
      uncertainty dominated first by ${\rm BR}(D\rightarrow\pi e\nu)$,
      then by $f_+$ \\
$D$ &  $K$  & $V_{cs}$ & uncertainty dominated by $f_+$     \\
$B$ &  $D$  & $V_{cb}$ & test corrections to heavy quark limit \\
$B$ & $\pi$ & $V_{ub}$ & vector final states useful; cf.\ text \\
\hline \hline
\end{tabular} \end{center}
\end{table}

The form factor is calculable.
The theoretical tools available are lattice QCD and symmetry arguments.
For example, chiral symmetry requires $f_+^{K\rightarrow\pi}(0)=1$,
with second-order corrections estimated to be $\lsim1\%$.
A combination of experimental measurements of the $q^2$ dependence with
this normalization condition gives the best determination of $V_{us}$
\cite{Leu84}.
It is not likely that lattice QCD will compete with this approach in
the foreseeable future, especially since the small quark masses in
$K\rightarrow\pi$ pose additional technical difficulties for the
lattice calculations.
Nevertheless, a comparison of the $q^2$ dependence of lattice and
experimental form factors, such as in Reference~\cite{BKS90}, could be used
to get a feel for the reliability of the quenched approximation.
Similarly, heavy-quark symmetry \cite{Isg90} requires
$f_+^{B\rightarrow D}(q_{\rm max}^2)=1$, again with second-order
corrections \cite{Luk90}.
Especially for the $D$-meson, the applicability of heavy-quark symmetry
is not guaranteed, but lattice QCD can be used to test it.

Lattice calculations of semi-leptonic form factors, essentially using
the strategy of Reference~\cite{BKS88}, have been carried out for
$D\rightarrow\pi,K$ \cite{BKS90,BKS88,Cri89,Aba90,Lub92} and
$D\rightarrow\rho,K^*$ \cite{BKS91,Lub91,BKS93}.
Two groups are involved, which we shall abbreviate
BKS \cite{BKS90,BKS88,BKS91,BKS93} and
ELC \cite{Cri89,Aba90,Lub92,Lub91}.
Both groups report statistical errors of roughly 15\%.
BKS also estimate systematic errors, which introduce an additional
30--40\% uncertainty; presumably the systematic uncertainties of the
ELC calculations are similarly large.
Except for the quenched approximation, however, the systematic
uncertainties would be smaller if the statistical errors were smaller.
For example, the largest contributor to the systematic uncertainty is
the lattice-spacing dependence of the form factors \cite{BKS90,BKS91}.
With better statistics over a wider range of lattice spacing, this
component can be reduced by extrapolating.

Table~\ref{semi-leptonic-results} summarizes lattice results for several
form factors in semi-leptonic decays of the $D$.
\begin{table}[t]
\caption[semi-leptonic-results]{Some results for form factors
$f_+(q^2)$ in $D\rightarrow K$ semi-leptonic decays and
$A_1(q^2)$ and $A_2(q^2)$ in $D\rightarrow K^*$ semi-leptonic decays.
Experimental results are from E691 and E653; their statistical and
systematic errors have been added in quadrature.
For BKS and ELC systematic errors are {\em not\/} listed.
Based on the estimates of BKS, it is reasonable to assign 30--40\%
systematic errors to form factors themselves and 20--30\% to the
ratio.}\label{semi-leptonic-results}
\begin{center}
\begin{tabular}{cr@{.}lr@{.}lr@{.}lr@{.}lr@{.}lr@{.}l}
\hline \hline
 &
\multicolumn{2}{c}{$f_+(0)$} &
\multicolumn{2}{c}{$f_+(q_{\rm max}^2)$} &
\multicolumn{2}{c}{$A_1(0)$} &
\multicolumn{2}{c}{$A_2(0)$} &
\multicolumn{2}{c}{$A_2/A_1(0)$} &
\multicolumn{2}{c}{$A_1(q_{\rm max}^2)$} \\
\hline
expt & 0&69(04) & \multicolumn{2}{c}{---} & 0&46(07) &
       \multicolumn{2}{c}{---} & ~0&82(25) & 0&54(08) \\
BKS  & 0&90(08) & 1&64(36) & 0&83(14) &
       0&59(14) & ~0&70(16) & 1&23(16) \\
ELC  & 0&63(08) & \multicolumn{2}{c}{---} & 0&53(03) &
       0&19(21) & \multicolumn{2}{c}{---} & \multicolumn{2}{c}{---} \\
%      0&19(21) & \multicolumn{2}{c}{---} & {\sl 1}&{\sl 10(26)} \\
\hline \hline
\end{tabular} \end{center}
\end{table}
Lattice results are most reliable at and near $q_{\rm max}^2$, i.e.\
when the spatial momentum of the hadrons is small.
However, especially by giving the initial-state meson non-zero momentum
\cite{Cri89}, it is possible to reach even $q^2<0$.
Experiments customarily quote results for the form factor at $q^2=0$,
so BKS and ELC do so too.
The extrapolation to $q^2=0$ is done by fitting to the pole-dominance
form
\begin{equation}
f_+(q^2)=\frac{f_+(0)}{1-q^2/m^2}
\end{equation}
where $m$ is a suitable resonance mass.
Both BKS and ELC find that their numerical calculations agree
qualitatively with this form.
However, verification of pole dominance is not essential to lattice QCD
or to experiments.
BKS stress the utility of a direct comparison for
vector-meson final states near the endpoint \cite{BKS93}.
Lattice calculations are most straightforward at $q_{\rm max}^2$, but
then vector modes are preferable experimentally, because the phase-space
suppression of the dominant form factor in the differential decay rate
at the endpoint is only $\lambda^{1/2}=2m_Ap_X$.

Although the uncertainty estimates on the results presented in
Table~\ref{semi-leptonic-results} are still at a qualitative stage, it
is important to realize that semi-leptonic decays are not much more
difficult to compute than the hadron masses and decay constants.
Since References~\cite{Wei92,Wei93} have demonstrated the feasibility of
a systematic, rather than incremental, approach, one can hope for a
comparable analysis of semi-leptonic decays in the near future.

\subsection{Neutral Meson Mixing}\label{mixing}
Some of the operators on the right-hand side of Equation~\ref{weak-OPE}
induce neutral meson mixing, e.g.\ $K^0\leftrightarrow\bar{K}^0$.
For the kaon the four-quark operator is
\begin{equation}\label{four-quark}
\cO_{\Delta S=2}=
\bar{s}_a\gamma_\mu(1-\gamma_5)d_a \bar{s}_b\gamma_\mu(1-\gamma_5)d_b ,
\end{equation}
where $a$ and $b$ denote color indices.
The mixing amplitude is proportional to the matrix element
$\langle\bar{K}^0|\cO_{\Delta S=2}|K^0\rangle$.
Similarly, the $\Delta B=2$ operator obtained by the substitution
$\bar{s}\mapsto\bar{b}$ induces $B^0$-$\bar{B}^0$ mixing.
On the other hand, $D^0$-$\bar{D}^0$ mixing is expected to be too small
to be interesting.

Let us focus on the kaon.
Phenomenologists use the so-called ``vacuum saturation approximation''
as a standard of comparison for
$\langle\bar{K}^0|\cO_{\Delta S=2}|K^0\rangle$.
This approximation treats the four-quark operator a product of
two-quark operators, inserts a complete set of states, and then keeps
only the vacuum contribution \cite{Gai55}.
The result is
\begin{equation}\label{VSA}
\left.\langle\bar{K}^0|\cO_{\Delta S=2}|K^0\rangle\right|_{\rm VSA}=
\frac{8}{3} m_K^2 f_K^2.
\end{equation}
The factor $8/3$ arises because there are two Fierz arrangements and
because both $\bar{s}$ operators can act on the initial state.
It is customary to define the ``kaon $B$ parameter''
\begin{equation}\label{B-K}
B_K=\frac{\langle\bar{K}^0|\cO_{\Delta S=2}|K^0\rangle}%
{\frac{8}{3}m_K^2 f_K^2}.
\end{equation}
In numerical lattice QCD the ratio $B_K$ is a convenient quantity,
because the statistical and systematic uncertainties of the ratio are
under better control than those of numerator or denominator separately.

A typical result using $B_K$ is the one for the parameter $\epsilon$,
which appears in the analysis of CP violation in the $K^0$-$\bar{K}^0$
system.
Combining the measurement of $|\epsilon|$ with other experimentally
known numbers, the standard model predicts
(cf.\ Reference~\cite{Nir92} and references therein)
\begin{equation}\label{K-epsilon}
\begin{array}{l}
5.6\times10^{-8} = \\[0.7em] \hspace{1.5em}
- \hat{B}_K\,|V_{cb}|\imag V_{td}
\left[\left(\eta_3f_3(y_t) - \eta_1\right)y_c|V_{cd}| +
\rule{0.0em}{0.97em}\eta_2y_tf_2(y_t)|V_{cb}|\real V_{td} \right],
\end{array}
\end{equation}
where $y_q=m_q^2/m_W^2$, $V$ is the CKM matrix, the $f_i$ are kinematic
functions, the $\eta_i$ are perturbative QCD corrections, and
$\hat{B}_K$ is a renormalization group invariant quantity related to
$B_K$.
Taking the one-loop anomalous dimension of $\cO_{\Delta S=2}$ into
account
\begin{equation}
\hat{B}_K=\left(\alpha_S^{}(\mu)\right)^{-2/9} B_K(\mu).
\end{equation}
The combination of CKM matrix elements in Equation~\ref{K-epsilon} depends
on the CP-violating phase and (using unitarity constraints) on
$|V_{ub}/V_{cb}|$.

As mentioned above, although there is no physical reason to prefer $B_K$
to the matrix element $\langle\bar{K}^0|\cO_{\Delta S=2}|K^0\rangle$,
it makes better sense to quote $B_K$ from lattice QCD.
Because of correlations in the Monte Carlo, the statistical fluctuations
of the numerator and denominator cancel to a large extent.
Moreover, an analysis based on chiral perturbation theory suggests that
some effects of the quenched approximation also cancel in the
ratio \cite{Kil90}.
Finally, $B_K$ should be finite in the chiral limit
($m_K^2\rightarrow0$), providing a consistency check
on the numerical results.

The most important reason why the lattice calculations of
$\langle\bar{K}^0|\cO_{\Delta S=2}|K^0\rangle$ are feasible is that
there can be no mixing with lower dimension operators, because
$\cO_{\Delta S=2}$ is the {\em lowest\/} dimension operator with
$\Delta S=2$.
Consequently, the numerical calculations presented below are much more
reliable than calculations of analogous matrix elements of penguins and
other denizens in the zoo of four-quark operators.
Despite these advantages, there are still some difficulties.
For Wilson fermions there are problems with chiral symmetry, making
necessary a subtraction \cite{Ber88,Gup92} that ultimately decreases
the signal-to-noise ratio.
For staggered fermions chiral symmetry makes this subtraction
unnecessary, but one must treat the extra flavors with care
\cite{Kil87}.

The numerical results with the smallest uncertainties have been done
with staggered fermions \cite{Kil90}.
At present the largest uncertainty comes from extrapolating in $a$; it
is uncertain whether the extrapolation should be taken in $a$ or $a^2$.
The most recent quenched results \cite{Sha92} are
$\hat{B}_K=0.66\pm0.06$ after a linear extrapolation and
$\hat{B}_K=0.79\pm0.03$ after a quadratic extrapolation.
By comparison, Wilson quarks yield $0.88\pm0.13$ \cite{Ber88,Gav88}.
A calculation in full QCD is compatible with the results from the
quenched approximation, supporting the arguments that effects of the
quenched approximation cancel in $B_K$ \cite{Kil92}.

These results for $B_K$ might foster the impression that the vacuum
saturation approximation gives a fair description, but that is
misleading.
Separating the four-quark operator into $VV$ and $AA$ terms, it turns
out that the two have large contributions that cancel in quenched
lattice QCD.
Conversely, the vacuum saturation approximation would assert that the
$AA$ term contributes everything and the $VV$ term nothing.

Mixing is also of great interest in the neutral $B$-meson system,
because, like $\epsilon$ in the neutral kaon system, it gives insight
into the third row of the CKM matrix.
In the standard model
\begin{equation}
x_d = (\mbox{known factors})\,|V_{td}^*V_{tb}|^2 f_B^2 B_B,
\end{equation}
where $x_d=\Delta M_{B^0}/\Gamma_{B^0}=0.66\pm0.11$ is a measure of the
mixing.
A similar formula applies to the $B_s$ meson.
In addition to the decay constant, discussed above, the $B$-meson $B$
parameter is needed.
Pilot lattice studies \cite{Aba92} yield values of $B_B$  and $B_{B_s}$
close to the vacuum saturation value of unity.
The level of technical detail in these calculations is not yet high
enough to understand all uncertainties, but a better
understanding will certainly emerge in the coming years.

\subsection{Non-Leptonic Decays} \label{non-leptonic}
Non-leptonic decays, such as $K\rightarrow\pi\pi$ processes, are also
mediated by four-quark operators from Equation~\ref{weak-OPE}.
Many of the interesting operators suffer from the problem of mixing with
lower dimension operators, which did not afflict the calculation of
$B_K$.
A more serious obstacle to the treatment of non-leptonic decays is the
presence of two (or more) hadrons in the final state.
The technical aspect is the difficulty of separating the particles in
the finite volume.
The conceptual aspect is the determination of final-state phase shifts
from purely real quantities computed in Euclidean field theories
\cite{Ber89,Mai90}.
It is rigorously known \cite{Lue91} how to determine the properties of
the $\rho$ resonance, which decays through an interaction in the QCD
Hamiltonian.
The stumbling block is evidently the application of the ideas in
Reference~\cite{Lue91} when the particle decays through an interaction being
treated as a perturbation, as for weak decays.
Note that these difficulties do {\em not\/} stem from the lattice
cutoff, but from other features, finite volume and imaginary time,
introduced to make the computational method tractable.
Nevertheless, until these issues are resolved, lattice results for
non-leptonic decays probably will not warrant attention from
non-experts.

\section{Summary and Prospects} \label{Summary}

% \paragraph{Beyond the quenched approximation?}
The coming generation of calculations will be done on computers with
speeds of tens of gigaflops.  In a few years, computers with hundreds of
gigaflops or perhaps a teraflop will probably be available \cite{Christ}.
These machines will make possible crucial improvements in lattice
calculations, but   increases in computing power alone
with no methodological improvements will probably
be insufficient to make possible
first principles calculations with light sea quarks.
Algorithms for the direct inclusion of sea quarks in QCD simulations made
dramatic progress during the 1980's.  The current best algorithms are orders
of magnitude more efficient than those proposed for the first Monte Carlo
spectrum calculations around 1980.
%\cite{WeingartenAlg}.
However, for large lattices and medium  quark masses they still exhibit
extremely long correlation times which are not understood theoretically,
and whose scaling behavior in such quantities as the lattice volume and
quark mass are not understood.
%Figure~\ref{fig:unquenched}
%\cite{Christcor}
The current consensus is that one order of magnitude in computing power is
likely to be
too little to do definitive calculations including light sea quarks, without
further theoretical insight.

% \paragraph{Living with the quenched approximation.}
What direction will lattice phenomenology take if there are no new
algorithmic ideas?
For heavy $Q\bar{Q}$ mesons, nonrelativistic arguments should make
possible rock solid understanding of all errors aside from quenching
errors, and decent understanding of those.
For hadrons containing light quarks, it now appears that good control of
all errors aside from quenching errors is likely  to be achievable in the
coming generation of calculations.
The uncertainties shown in Figure~\ref{fig:spectrum} \cite{Wei92},
will be checked in the coming few years (and perhaps reduced to the point
that the degree of disagreement of the quenched approximation with the real
world stands out more clearly).
If the mass of one of the light hadrons were unknown, one might take the
typical disagreement with the known quantities as a phenomenological estimate
of the quenching uncertainties.

% \paragraph{Summary}
Today there are a couple of phenomenologically interesting lattice QCD
calculations in which, because they are in one way or another special cases,
a complete error analysis has been attempted.
For the more demanding case of the light hadron spectrum,
a systematic calculation this year made corrections for all of the
sources of error within the quenched approximation, but did not completely
estimate the uncertainties of all of the corrections.
It is likely that such estimates for light hadron calculations
will prove possible in the quenched approximation on the current and
coming generation
of large scale computers without great conceptual breakthroughs (although
with much labor).
If this becomes the case, many of the most crucial applications
of lattice QCD to standard model phenomenogy
will be likewise calculable.

% \section*{Acknowledgements}
% We thank EE, AXK, GPL, DW.
%\section*{References}

\end{document}